\documentclass[aps,amsmath,amssymb,superscriptaddress,twocolumn,showkeys,showpacs,letterpaper]{revtex4}
\usepackage{textcomp}
\usepackage{graphicx}
\usepackage{hyperref}
\def\ten#1{\oalign{$\bf #1$\crcr\hidewidth$\scriptscriptstyle\sim$\hidewidth}\vphantom{#1}}

\begin{document}

\title{Finite element analysis of fretting crack propagation}

\author{H. Proudhon}
\thanks{Corresponding author: henry.proudhon@mines-paristech.fr.}
\affiliation{MINES ParisTech, Centre des Mat\'eriaux, CNRS UMR 7633, BP 87 91003 Evry Cedex, France}
\author{S. Basseville}
\affiliation{LISV, Universit\'e de Versailles Saint-Quentin, 45 avenue des Etats-Unis, 78000 Versailles, France}

\begin{abstract}
In this work, the finite elements method (FEM) is used to analyse the growth of fretting cracks. FEM can be favourably used to extract the stress intensity factors in mixed mode, a typical situation for cracks growing in the vicinity of a fretting contact. The present study is limited to straight cracks which is a simple system chosen to develop and validate the FEM analysis. The FEM model is tested and validated against popular weight functions for straight cracks perpendicular to the surface. The model is then used to study fretting crack growth and understand the effect of key parameters such as the crack angle and the friction between crack faces. Predictions achieved by this analysis match the essential features of former experimental fretting results, in particular the average crack arrest length can be predicted accurately.
\end{abstract}


\maketitle
\section{Introduction}
\label{intro:sec}
Fretting damage is still a major issue for a large number of multi-parts industrial applications subjected to vibrations \cite{Waterhouse1992}. Many palliatives have been developed over the years such as modifying the geometry, coating surfaces to reduce friction or using shot penning in some cases \cite{Fouvry2006}. Very small oscillatory displacement of surfaces in contact induce partial slip conditions which can lead to rapid crack initiation and imped the life of assembled structures. Well known examples can be blade/disk TA6V contact in jet engines or Al alloys riveted lap joints in aerospace structures.

Various models have been developed over the years \cite{Szolwinski1998, Madge2008} to predict the growth of fretting fatigue cracks but they usually fail to predict the early stages of the crack growth (typically before 1 mm). This stage is termed short crack behaviour and usually exhibits strong deviations with respect to the long crack domain which can be attributed to a combination of mixed mode growth under the influence of the highly multi-axial stress field, plasticity and roughness induced crack closure and crystallographic effects \cite{McDowell1996, Doquet1999, Ludwig2003}.

Recent developments in short fatigue crack models tend to take into account more and more of those ingredients, but fretting crack growth is very often still described at the macroscopic level. Notable exception are recent studies which take into account the grain microstructure in the contact region but they are limited to the initiation of fretting cracks \cite{Neu2008, Dick2008} and a study of short fretting crack propagation using dislocation distributions \cite{Fouvry2008}.
This work is a first step to study how FEM analysis can be favourably used to access local material behaviour at the crack tip to predict the overall fretting crack growth. We focus on the simple case of straight cracks initiated at the contact border; those cracks typically evolve in a multiaxial stress state which will strongly depend on the angle of propagation. Previous study of fretting crack propagation usually neglect mode II and use the assumption of Elber ($\Delta K=K_{max}$) to determine the stress intensity factor range \cite{Fadag2008, Giner2008}. While this might be justified for a fatigue crack under uniaxial load with a negative stress ratio, the case of fretting loading is much more complicated and there is no obvious reason for this assumptions to stand. In particular, the friction between the crack faces may influence mode II stress intensity levels due depending on the crack angle \cite{Ribeaucourt2007}.

First the FE fretting model with straight cracks is described and two popular weight functions are used to test and validate the stress intensity levels in the case of cracks growing perpendicularly to the surface. FE computations are used to predict the mixed mode stress intensity levels during fretting cycles (thus accessing both $K_{max}$ and $K_{min}$) and to determine the influence of key parameters such as the crack angle and the effect of friction between crack faces. Stress intensity ranges are eventually combined to the Paris law to predict crack growth and arrest levels. Those predictions are compared to experimental crack lengths observed on interupted tests in an 2xxx aluminium alloy.

\begin{figure}[!htb]
 \centering
 \includegraphics[width=85mm]{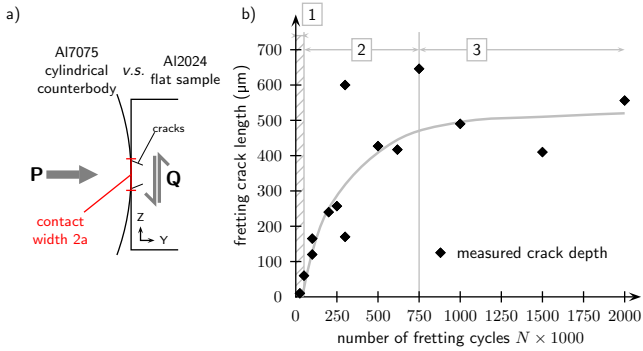}
 \label{fretting_test_and_propagation:fig}
 \caption{a) Simplified schematics of a fretting wear test in cylinder plane configuration, b) Measured fretting crack depth as a function of the number of cycles in a 2xxx series aluminium alloy, $P=440$~N/mm and $Q=240$~N/mm; the three domains labelled 1,2 and 3 correspond respectively to crack initiation, fast propagation and crack arrest, after~\cite{Munoz2006}.}
\end{figure}

\section{Numerical Methods}
\label{methods:sec}

\subsection{Fretting model}
\label{fretting_model:ssc}
The simulations ultimately aim at predicting real fretting test and thus have been designed to closely match the experimental geometry, materials and loading conditions.

The cylinder/plane fretting wear test shown in Fig.~\ref{fretting_test_and_propagation:fig}a) is represented in a 2D plane strain FE computation using the finite element code Z-SeT/Z\'eBu\-LoN \cite{Besson1998}. Far from the contact zone, free meshing with plane strain 3-node triangular full integration elements is used and the mesh is gradually refined towards the contact region. To combine a high precision with reasonable computation times, the mesh in the contact zone is paved with plain strain rectangular 4-node full integration elements with a size of $20$~\textmu m. With the investigated conditions, this provides a contact element over contact area ratio of 80 which is sufficient to accurately describe the multiaxial field imposed to the sample by the contact conditions (For a specimen free of cracks, the stresses computed with this geometry fits the analytical solution within 2\%).

The loading is applied in two steps: first, a normal displacement $u_2$ is imposed on the top of the half cylinder to cause a normal reaction of 410~N/mm (corresponding to a half contact width of $a=0.8$~mm and a Hertz pressure of 325~MPa). The plate is kept fixed by locking its bottom and side nodes.
An horizontal cyclic displacement $u_1$ is imposed to the bottom and side nodes of the plate to cause a tangential reaction of 240~N/mm.

For the present study the material is described by isotropic elasticity; elastic constants $E=72$~GPa and $\nu=0.3$ are used to simulate a 2xxx aluminium alloy. Contact between the two parts is modelled by a classical impactor/target technique and solved with a flexibility method implemented in Z-SeT/Z\'eBuLoN. Technically, the contact reactions are at first computed in a local contact algorithm and then added to the global problem. The friction behaviour is introduced by the Coulomb's law with a friction coefficient of 1.1, which represents the friction behaviour of a dry aluminium/aluminium contact~\cite{Proudhon2005}.

\subsection{Finite element analysis of fretting cracks}
\label{fea_crack:ssc}

In this approach, the cracks are explicitly introduced within the FE mesh. A in-house program is used to mesh both cracks independently, using as input their geometry (here initial position of the crack, length and angle). Two rectangular spaces are reserved on both side of the contact area to receive the cracks. While both crack faces are initially superimposed, they are can open and close during the computation and the contact conditions between them have to be defined. The crack tip is meshed with a fine, regular region to ensure a good description of the stress singularity; this region is used to compute the energy release rate $G$. All these aspects are schematically summarised in Fig.~2.
\begin{figure}[!htb]
 \centering
 \includegraphics[width=85mm]{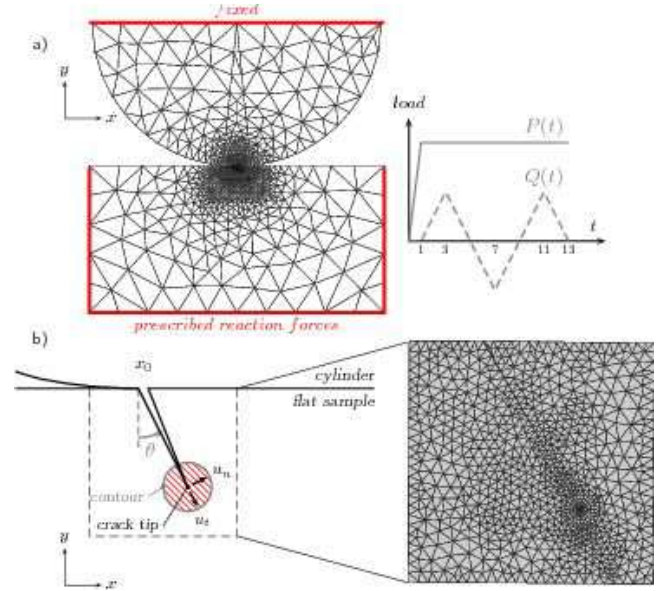}
 \label{fem_contact_zoom_left_crack:fig}
 \caption{Cracks are explicitly introduced within the finite elements model with a refined mesh at the crack tip to compute the energy release rate $G$.}
\end{figure}

Once cracks have been introduced within the mesh, finite element computation can be carried out to determine the stress distributions in the neighbourhood of the crack tip during the fretting cycle. Due to the cyclic reci\-procating motion of the counter body, both crack will alternatively open and close. As an example, Fig.~3 shows the mode I stress intensity factor for the left crack during one fretting cycle. As expected the left crack opens when the counterbody is moved to the right. By computing a complete fretting cycle, the stress intensity factor ranges $\Delta K_I$ and $\Delta K_{II}$ can be determined.

\begin{figure}[!htb]
 \centering
 \includegraphics[width=85mm]{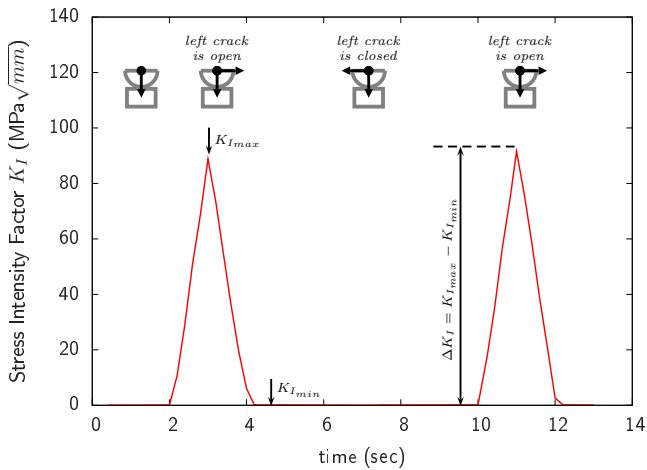}
 \label{K_cycle_7_straight_15:fig}
 \caption{Example of left crack stress intensity factor extracted form the FEM calculation; here the crack is straight, 400 \textmu m long and inclined towards the center of the contact ($\theta=15$\textdegree).}
\end{figure}

This analysis is used iteratively to compute the stress intensity levels during the growth of fretting cracks. Each different stage uses a different mesh accounting for the appropriate crack length and angle. The complete analysis is as follows:
\begin{enumerate}
\item For the first stage, the crack initiation position $x_0$ and initial angle $\theta_0$ have to be defined. This can be done using a fatigue multi-axial criterion (such as Dang Van or SWT) as discussed thouroughly by many authors (see~\cite{Fouvry2002} for a review). The present work focuses on propagation and those values are chosen somewhat arbitrarily. $x_0$ is chosen equal to $0.96 a$ and $\theta_0$ will vary from zero (crack is perpendicular to the surface) to 45\textdegree{} (crack is inclined towards the center of the contact).
\item The cracks are introduced within the mesh as explained earlier and a complete fretting cycle is computed.
\item The energy release rate $G$ at the crack tip is determined by post-processing the finite element results with a perturbation method~\cite{Parks1974}.
\item Local displacements fields $u_n$ and $u_t$ respectively normal and tangent to the crack tip direction are computed and the ratio $q=u_t/u_n$ is evaluated over the 4 pair of nodes behind the tip (this correspond to a 20~\textmu m distance).
\item Both mode I and mode II stress intensity factors are deduced from \[q=K_{II}/K_I\] and
\[G=\frac{K_I^2}{E^\prime}+\frac{K_{II}^2}{E^\prime}\] with $E^\prime=E/(1-\nu^2)$ in plane strain.
\item The crack geometry is updated with a new segment $(l_i,\theta_i)$, $\theta_i$ being kept constant  and $l_0$ being equal to 25~\textmu m.
\end{enumerate}

\subsection{Weight functions for straight cracks perpendicular to the surface}
\label{weight_functions:ssc}

It is worth noting that thanks to R. D. Mindlin, the analytical solution of the partial slip regime is known precisely \cite{Mindlin1953}. The analysis was derived for sphere/sphere contact but can be used also for a 2D cylinder/plane contact and provide the stress state of any material point $M(x,y)$ of the plane during the complete fretting cycle ($P$ held constant, $Q(t)$ varying from $+Q$ to $-Q$).

For very small displacement amplitudes, slip occurs only on the outer edges of the contact, the central zone $|x|<c$ remaining in stick condition. During the fretting cycle, the tangential force goes from $+Q$ to $-Q$ and the stick/slip limit $c'$ evolves from $c$ to $a$. The partial slip contact can be described during the fretting cycle, by superimosing the elastic solution from the normal load only $\ten\sigma^n$ and three slipping contact elastic solutions $\ten\sigma^t$: (i) a slipping contact between $-a$ and $+a$ with a slip amplitude of $+\delta$ (ii) a slipping contact between $-c$ and $+c$ with a slip amplitude of $+\delta$ (iii) a slipping contact between $-c'$ and $+c'$ with a slip amplitude of $-2\delta$.

Using this description, the stress state can be derived for any material point within the plane for a given fretting loading ($P$,$Q$). One can then describe the stress evolution along a virtual crack path during the fretting cycle \cite{Hills1994}. In particular, at the end of the fretting cycle $c'=c$ the stress tensor at the position normalized by the conact width $(x/a,y/a)$ is:
\begin{equation}
\ten \sigma\left(\frac xa,\frac ya\right)
  =\underbrace{\ten \sigma^n\left(\frac xa,\frac ya\right)}_{\scriptsize\textrm{norm. load}}
  +\underbrace{\ten \sigma^t\left(\frac xa,\frac ya\right)}_{\scriptsize\textrm{tang. forward load}}
  -\underbrace{\frac ca \ten \sigma^t\left(\frac xc,\frac yc\right)}_{\scriptsize\textrm{tang. reverse load}}
\label{PSR_stress.eqn}
\end{equation}
Stress from Eq.~(\ref{PSR_stress.eqn}) can be favourably used as input for a weight function of a straight crack to predict the stress intensity factor of fretting cracks.

Many weight functions have been developed over the years since the seminal work of Bueckner \cite{Bueckner1970}. In this work, two weight functions for a single edge crack in a plate will be used to validate the finite element approach in the case of a straight crack perpendicular to the surface. The first one is Bueckner weight function and the second one is an approximated weight function from Wu and Carlsson which account for the finite width of the plate \cite{Wu1991}. The stress intensity factor is expressed as \[K=f\sqrt{\pi aW}\]
$a$ is the crack length, $W$ the specimen thickness and
\begin{equation}f=\sqrt{W}\int_0^a\sigma(x)\,m(a,x)\,dx
\label{wf:eqn}
\end{equation}
with $\sigma(x)$ being the stress normal to the crack face along the crack path evaluated in the non cracked specimen from Eq.~(\ref{PSR_stress.eqn}) and \[m(a,x)=\frac{1}{2\pi a}\sum_{i=1}^5\beta_i(a)\left(1-\frac{x}{a}\right)^{i-3/2}\] where the $\beta_i(a)$ functions can be found in \cite{Wu1991}.

\section{Results}
\label{results:sec}

The model presented in section~\ref{methods:sec} has been used to study the influence of the crack path amid the complex fretting stress field. First, straight cracks perpendicular to the surface are investigated and weight functions are used to compare with values extracted from finite elements calculations. Then the influence of the crack angle with respect to the surface on the stress intensity factors range evolution as the crack propagates is regarded; the effect of friction on crack faces is also studied. Eventually life predictions are derived from the stress intensity factor ranges and compared to experimental values.

\subsection{Straight crack perpendicular to the surface}
\label{straight_crack:ssc}

\begin{figure}[!htb]
 \centering
 \includegraphics[width=85mm]{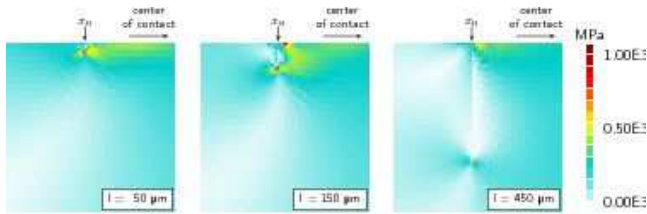}
 \caption{Isovalues of von Mises stress for 3 different crack lengths: 50, 150 and 450~\textmu m; straight crack perpendicular to the surface.}
 \label{vm_straight_00:fig}
\end{figure}

Cracks are introduced in the mesh at $x_0=\pm0.77$. Both stress intensity factor ranges are extracted at each time increment during the fretting cycle. After each calculation (starting with a 25~\textmu m crack), the crack length is increased by 25~\textmu m. Selected results are shown on Fig.~\ref{vm_straight_00:fig}.

\begin{figure}[!htb]
 \centering
 \includegraphics[width=85mm]{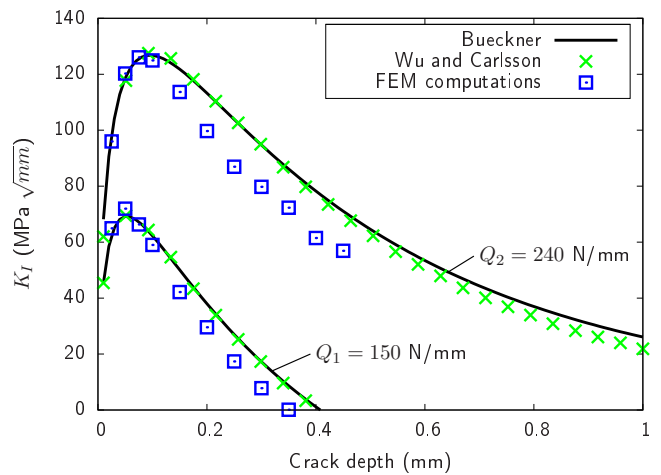}
 \caption{Comparison of mode I stress intensity factors for a straight fretting crack perpendicular to the surface computed by means of weight functions and finite elements.}
 \label{KI_perp:fig}
\end{figure}

Computations have been carried out for two different values of the tangential force ($Q1=150$ N/mm and $Q2=240$ N/mm). The maximum stress intensity factor for mode I has been extracted as explained in section\ref{fea_crack:ssc} and are compared to the values predicted by the use of weight functions (see section~\ref{weight_functions:ssc}). For the case of cracks perpendicular to the surface, normal loading of the crack face is $\sigma_{xx}(y)$ described by which is obtained analytically from Mindlin analysis and used in Eq. (\ref{wf:eqn}) to derive the evolution of $K_I$ as a function of the crack length. For small crack lengths, the values computed by FE are in very good agreement with those calculated with weight functions as for longer crack lengths, the value computed by FE is slightly lower. Presumably this can be attributed to the finite length of the sample which is not taken into account in the weight functions used here; a similar effect can been observed in Single Edeg Notched Tensile (SENT) specimens under uniaxial loading. Indeed, computing $K_I$ by Bueckner's weigth function deviates from the FE values when the crack size increases. For the SENT case, a solution accounting for the finite size of the specimen \cite{Ahmad1991} can be used which match the FE prediction very well. Nevertheless this approach validates the FE computations and the stress intensity levels predicted by this method. Regardless of the tangential force value, $K_I$ first increases with crack length, reaches a maximum around 100 \textmu m and then drops progressively. The drop in crack driving force was expected since with pure fretting conditions there is no bulk load and the stress level decreases when moving away from the contact area. The effect of the tangential force magnitude is also clearly highlighted by Fig.~\ref{KI_perp:fig}, increasing $Q$ lead to an increase in maximum crack driving force. The magnitude of $Q$ also seems to influence the drop of $K_I$, and one can qualitatively predict a longer crack with a higher $Q$ which is in agreement with experimental observations.

\subsection{Slanted crack}
\label{slanted_crack:ssc}

\begin{figure}[!htb]
 \centering
 \includegraphics[width=85mm]{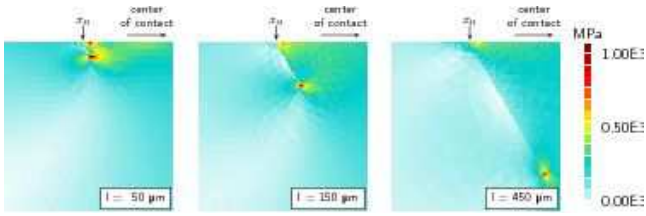}
 \caption{Isovalues of von Mises stress for 3 different crack lengths: 50, 150 and 450~\textmu m; straight crack inclined with a 30\textdegree{} angle.}
 \label{vm_straight_crack:fig}
\end{figure}

Experimental observations of fretting cracks by metallography techniques often show that cracks can be inclined towards the center of the contact. The angle is expected to have an influence on the mode I and mode II levels due to the multiaxial stress state in the subsurface contact region. FE computations series have been carried out with several values of the crack angle ranging from 15\textdegree{} to 45\textdegree (0\textdegree{} being perpendicular to the specimen surface). Selected results are shown on Fig.~\ref{vm_straight_crack:fig} for the cases of 30\textdegree.

\begin{figure}[!htb]
 \centering
 \includegraphics[width=42mm]{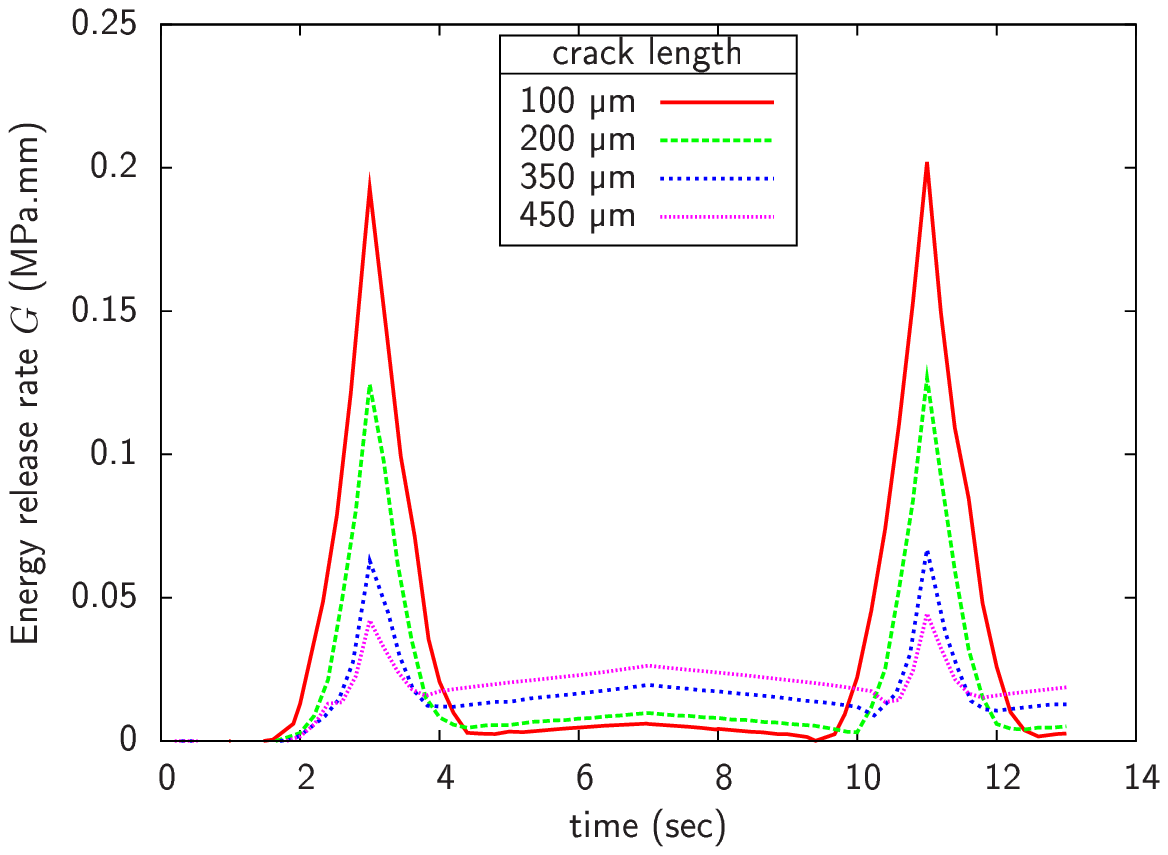}
 \hfill
 \includegraphics[width=42mm]{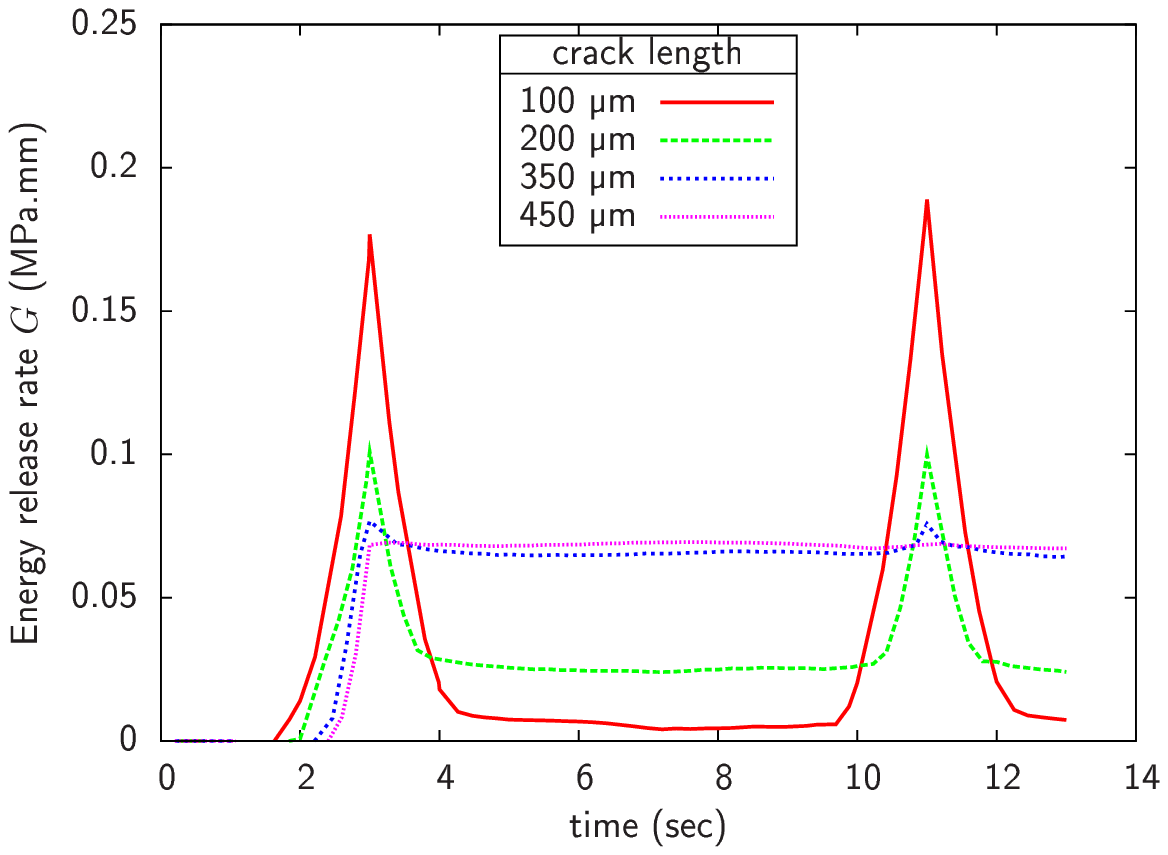}
 \newline
 \includegraphics[width=42mm]{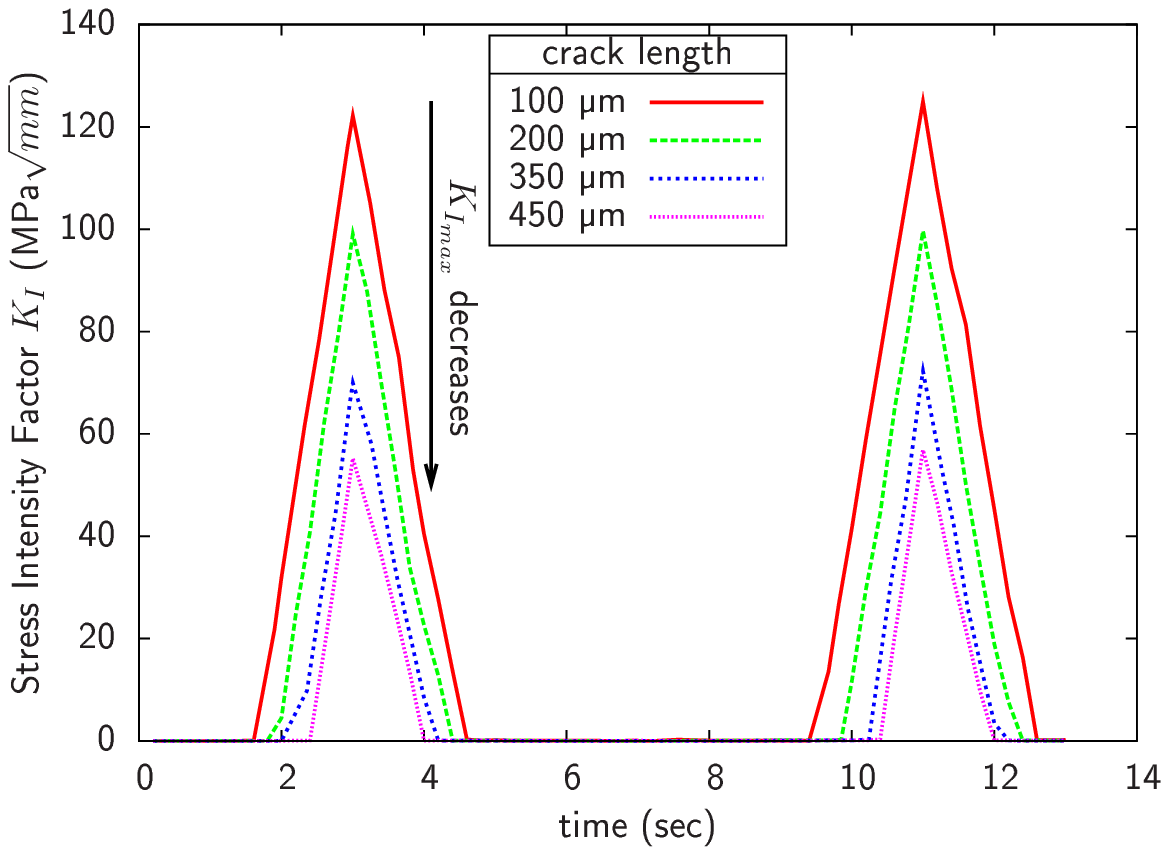}
 \hfill
 \includegraphics[width=42mm]{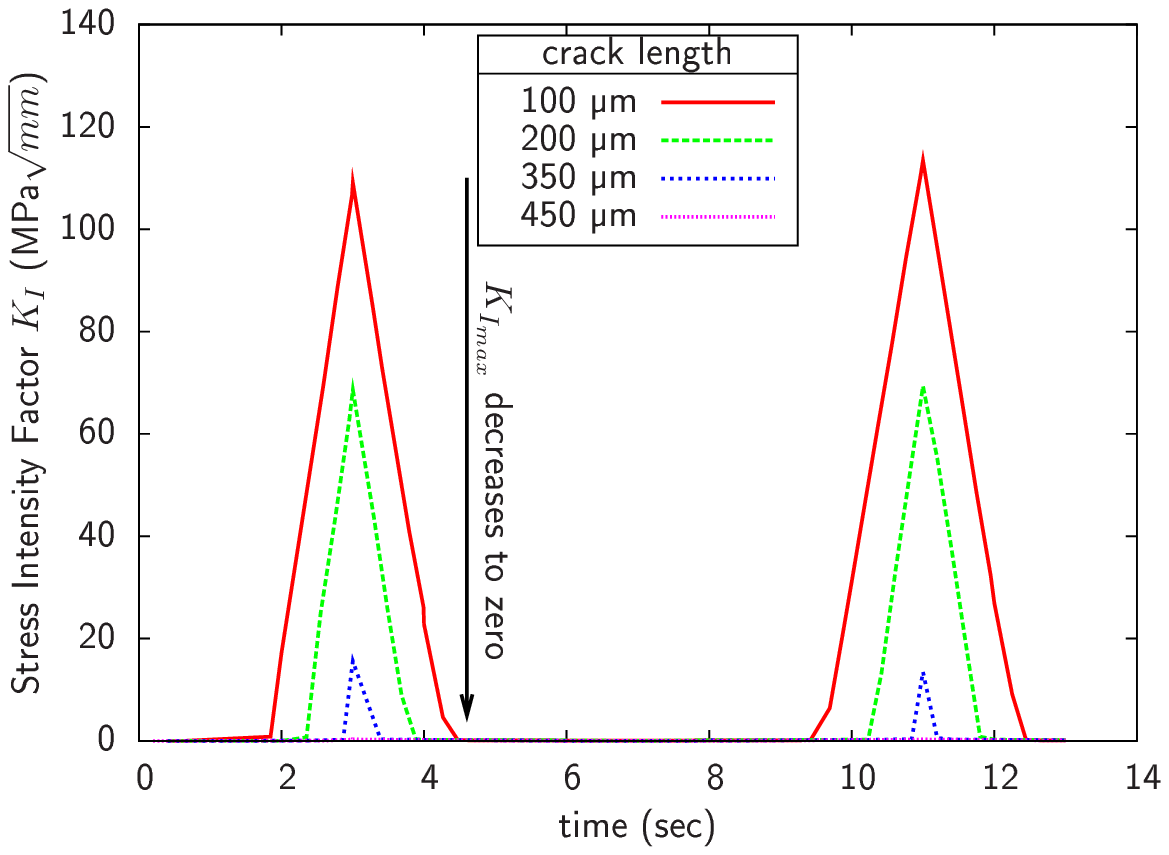}
 \newline
 \includegraphics[width=42mm]{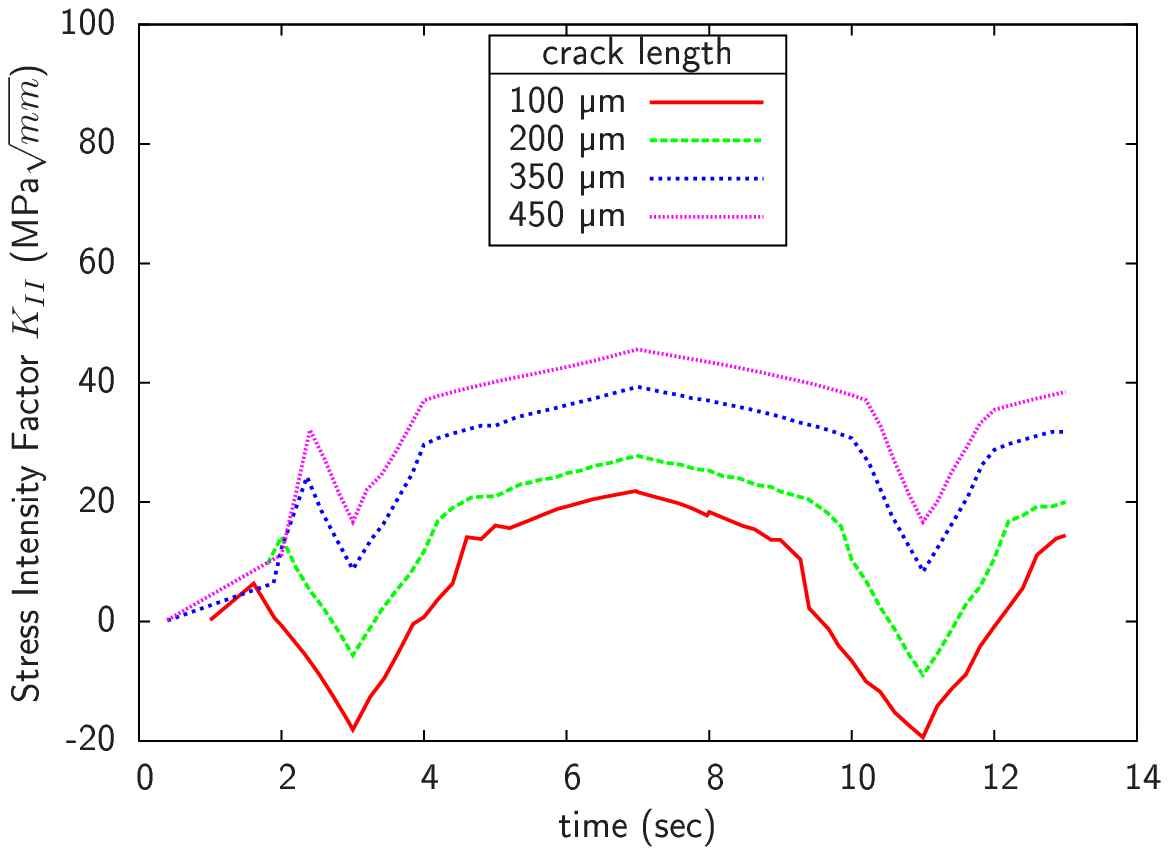}
 \hfill
 \includegraphics[width=42mm]{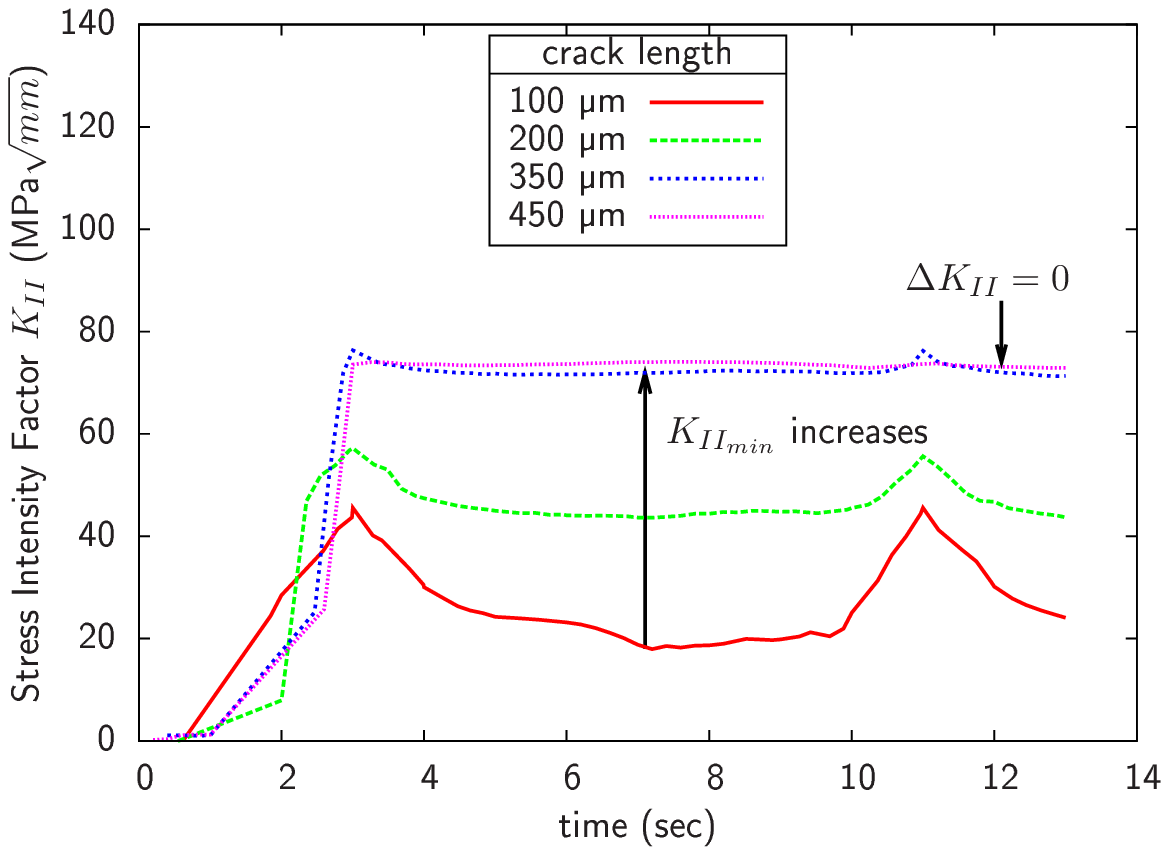}
 \caption{Energy release rate $G$ and stress intensity factors $K_I$ and $K_{II}$ evaluated during the fretting cycle: left column shows results for a straight crack perpendicular to the surface whereas right column shows results for an inclined crack ($\theta=30$\textdegree); period of the fretting cycle is 8 seconds.}
 \label{K_cycle_straight:fig}
\end{figure}

The stress intensity factors have been computed as explained in section~\ref{fea_crack:ssc} for a variety of crack angles and crack lengths. A comparison for the cases of 0 and 30\textdegree{} is given in Fig.~\ref{K_cycle_straight:fig} for different crack lengths.

\begin{figure}[!htb]
 \centering
 \includegraphics[width=0.48\textwidth]{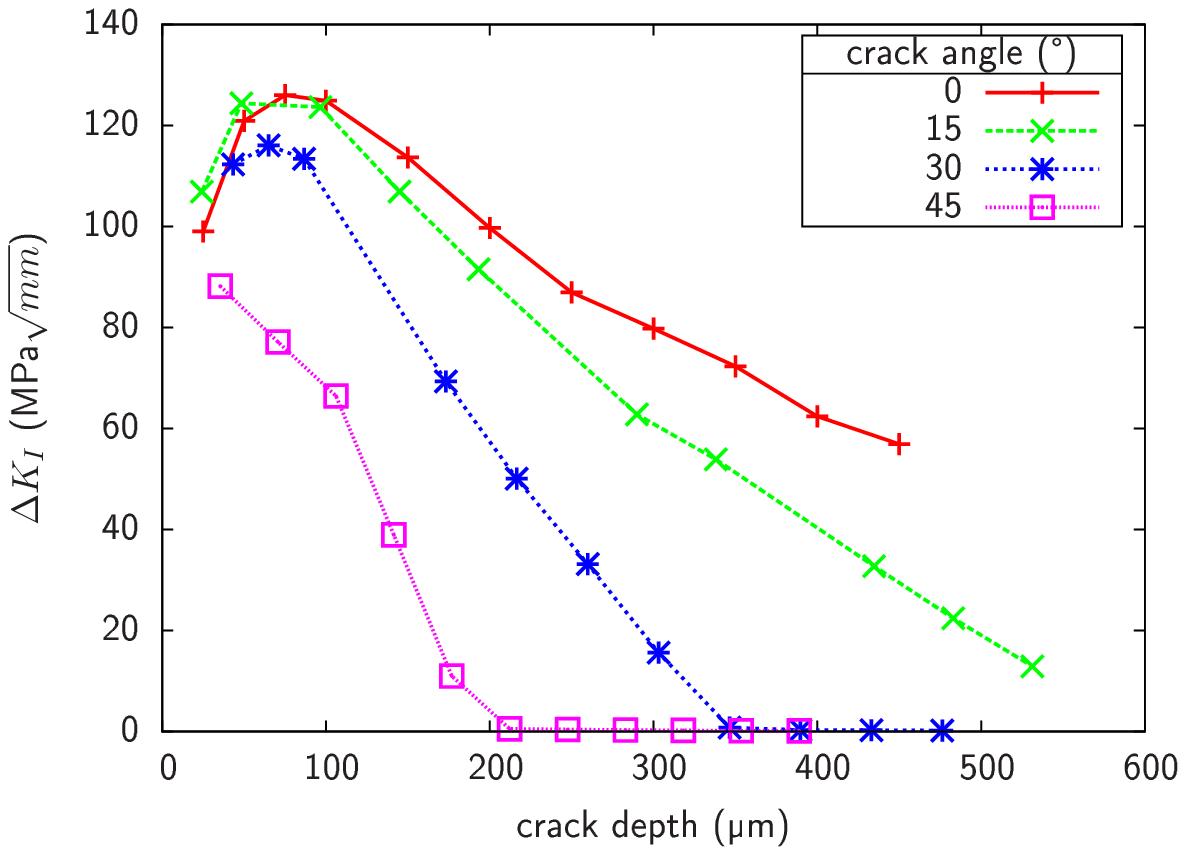}
 \hfill
 \includegraphics[width=0.48\textwidth]{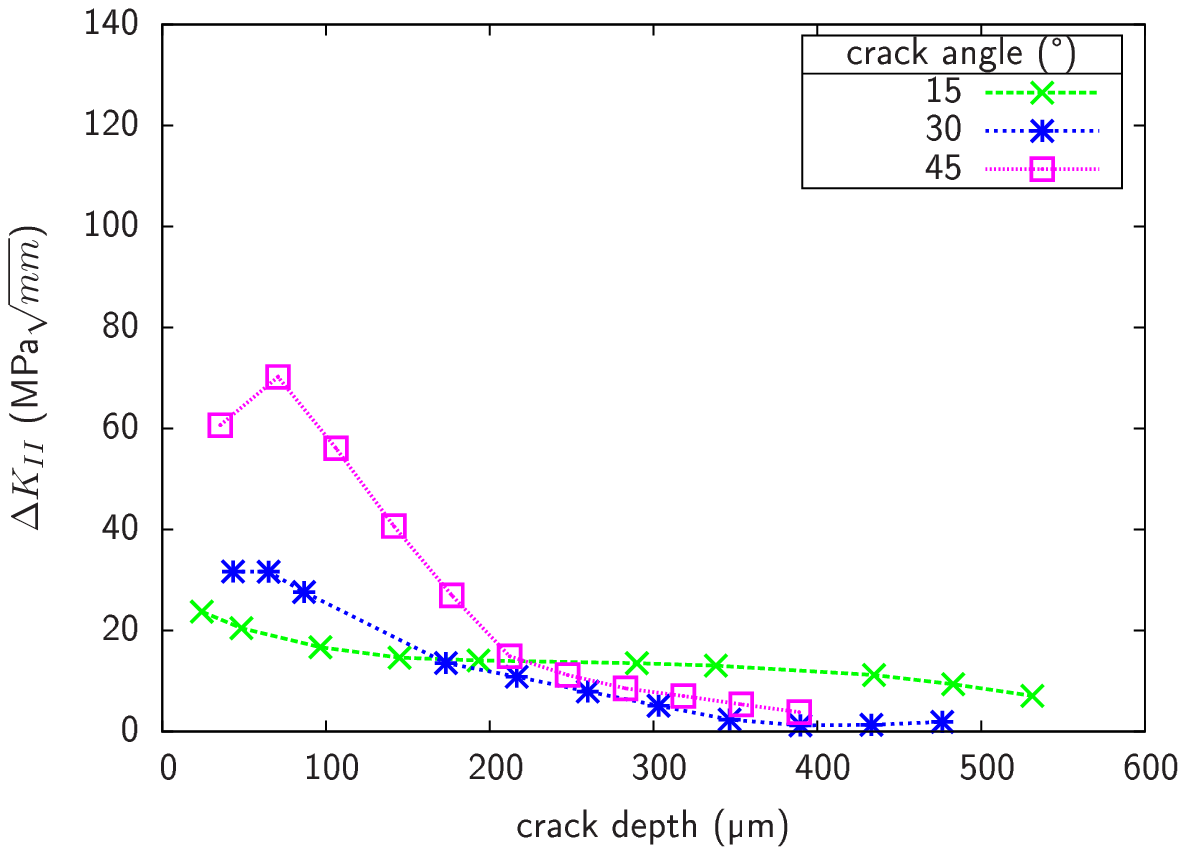}
 \caption{Evolution of the stress intensity factor ranges $\Delta K_I$ and $\Delta K_{II}$ as a function of crack depth for all investigated crack angles.}
 \label{summary_delta_K:fig}
\end{figure}

Fig.~\ref{K_cycle_straight:fig} highlights the importance of considering the whole fretting cycle. For $K_I$ levels, the maximum is always reached at the end of the fretting cycle, where the crack is open. For cracks perpendicular to the surface, $K_{I_{min}}$ remains close to zero and the decrease in terms of stress intensity factor range is due to the decrease of $K_{I_{max}}$. For slanted cracks, the drop of $\Delta K_I$ as a function of crack depth is more pronounced due to the crack angle. The more the crack is oriented towards the center of the contact, the faster the crack driving force will drop. Considering $K_{II}$ levels, the maximum occurs at a different time during the fretting cycle for crack perpendicular to the surface and slanted cracks. It is interesting to see that for slanted cracks, $K_{II_{max}}$ values will not decrease to zero as for $K_{I_{max}}$, but increase. However, $K_{I_{min}}$ also increases, resulting in a decrease of $\Delta K_{II}$ levels as a function of crack depth are shown on Fig.~\ref{summary_delta_K:fig}. We shall demonstrate in the next section that this can be predominantly attributed to the friction of the crack faces.

Fig.~\ref{summary_delta_K:fig} shows a summary of the evolution of the stress intensity factor range $\Delta K_I$ as a function of crack depth for all investigated crack angles (0, 15, 30 and 45\textdegree).

\subsection{Effect of friction on crack faces}
\label{friction_effect:ssc}

\begin{figure}[!htb]
 \centering
 \includegraphics[width=0.48\textwidth]{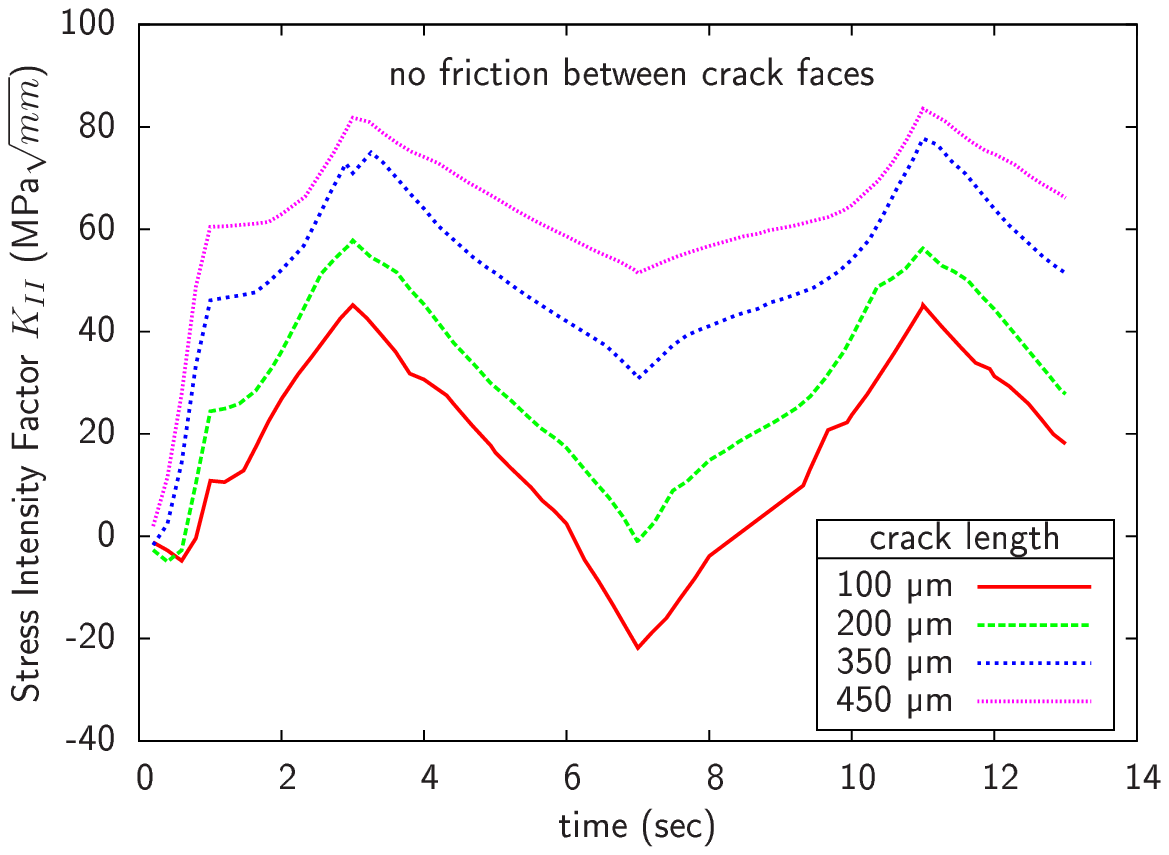}
 \hfill
 \includegraphics[width=0.48\textwidth]{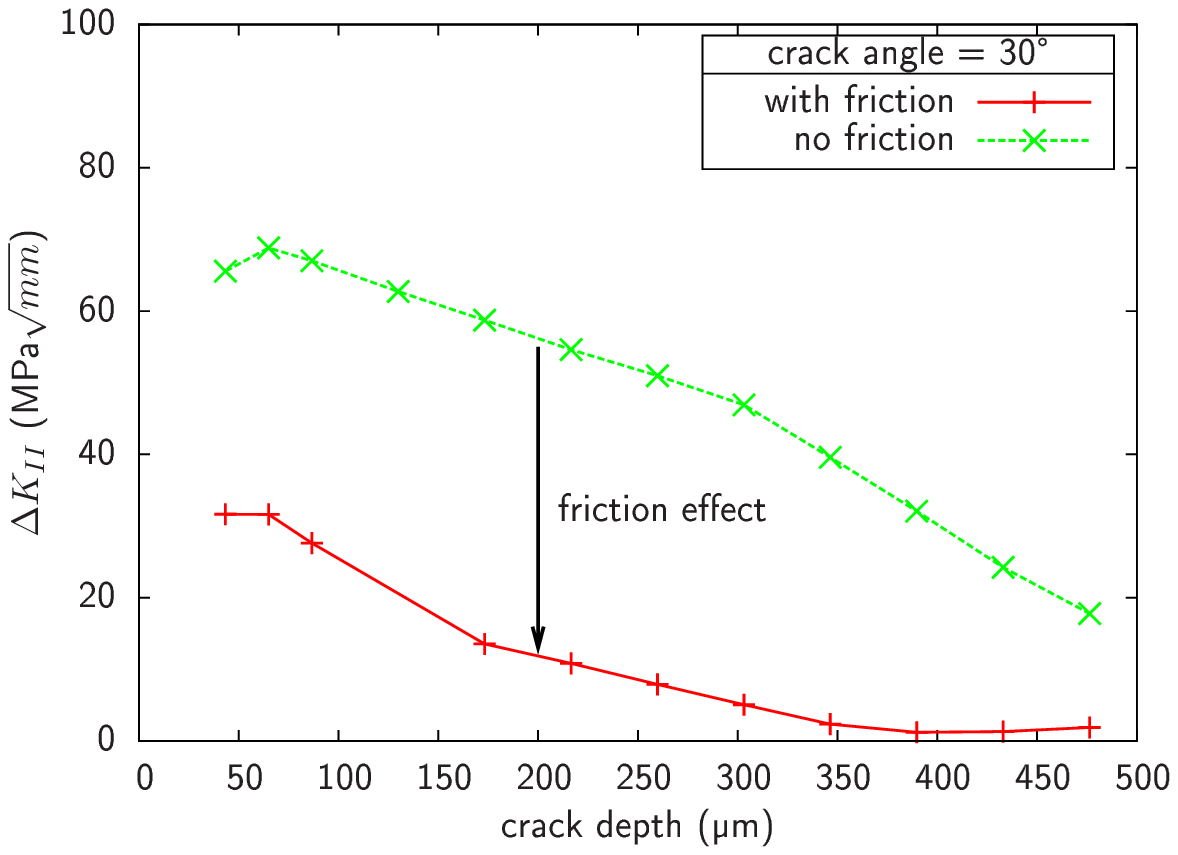}
 \caption{Effect of the friction between crack faces on mode II stress intensity levels.}
 \label{straight_30_mu:fig}
\end{figure}

To check the influence of friction applied between the crack faces, computations for the crack inclined with a 30\textdegree angle has been carried out a second time enforcing zero friction between the crack faces. Results of these calculations are depicted in Fig.~\ref{straight_30_mu:fig}. This does not change the maximum values of the stress intensity factors, since those are obtained at the end of the fretting cycle when the crack is open. However, this clearly shows that the minimum value of $K_{II}$ is significantly lower leading to higher $\Delta K_{II}$ levels. This is due to the fact that slanted cracks evolve in mixed mode and friction will limit crack tip shear displacement. If friction of crack lips is enforced, stick condition will be reached at some point preventing the shear displacement of crack tip nodes and leading to a $K_{{II}_{min}}$ value close to $K_{{II}_{max}}$ as seen on Fig.~\ref{K_cycle_straight:fig} for a crack larger than 350 \textmu m. This result will strongly impact life predictions (maximum crack depth as well as number of cycles to reach a certain depth) when considering mixed mode crack growth.

Those results highlight the key role of the friction between crack faces. This effect is particularly expected in the absence of bulk fatigue load (in pure fretting conditions) where the mode I stress intensity factors will progressively decrease to zero and mode II will become predominant. The crack progression will then strongly be controlled by the friction between the crack faces. A high friction will prevent shear tip displacement and stop the crack. High observed friction in experiments could also be related to roughness induced crack closure, since real fretting cracks may exhibit either strong deviations due to a crystallographic crack path, either small roughness as observed by post mortem examination or X-ray tomography slices \cite{Proudhon2007}. In both case, RICC levels are expected to be quite high which would lead to lower $\Delta K$ levels and a lower crack arrest depth~\cite{Parry2000b}.

\subsection{Life Predictions}
\label{life:sec}
Using the previous calculations, life predictions based on the computed stress intensity factor range --or more exactly crack depth prediction as a function of the number of cycle, since specimens do not fail in pure fretting conditions-- can be derived from the Paris Law: \[N=N_i+\int_{l_0}^{l_n}\frac{1}{C\Delta K_{eff}^m}da\] where $K_{eff}=\sqrt{K_I^2+K_{II}^2}$, the coefficient $C$ and the exponent $m$, are given for $da/dN$ expressed in mm/cycle and $K$ in MPa$\sqrt{m}$. This equation has been originally developed for mode I fatigue crack growth but can also reasonably well describe mixed mode crack growth as shown in \cite{Lamacq1996}. The two constants are equal to $C= 6.53 \times 10^{-8}$mm/cycle, $m= 3.387$ for the considered 2xxx aluminium alloy. For the studied fretting conditions, $N_i$ has been experimentally determined to be equal to $50.10^3$ cycles~\cite{Proudhon2005}.

\begin{figure}[!htb]
 \centering
 \includegraphics[width=0.48\textwidth]{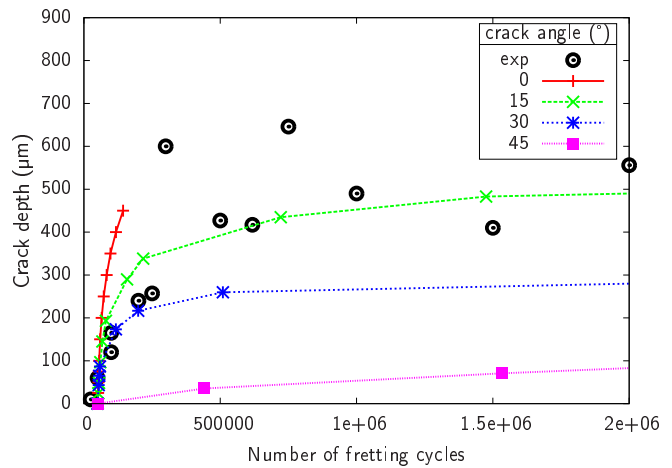}
 \caption{Life predictions based on the stress intensity factors determined by FEM for different crack angles; crack depth measured experimentally have been added for comparison.}
 \label{life_straight:fig}
\end{figure}

Fig.~\ref{life_straight:fig} shows the results of the fretting life predictions for the same loading conditions as used in the experiments \cite{Munoz2006}. One can readily observe that the crack angle has a strong influence on the propagation life as suggested by the $\Delta K_I$ and $\Delta K_{II}$ values already presented. Increasing the angle of the crack towards the center of the contact will have two effect: (i) the crack growth will be slower and (ii) the crack arrest depth will be shorter. For clarity, predictions for the 30\textdegree{} angle without friction between the crack faces are not presented in Fig.~\ref{life_straight:fig} but the curve is essentially the same with a longer crack arrest depth (300 \textmu m versus 200 \textmu m in this particular case). This is clearly due to the higher $\Delta K_{II}$ value which rapidly becomes predominant for inclined cracks ($\Delta K_I$ reaches zero).

Regarding the comparison with the experimental results, one can observe that the 30\textdegree{} angle predictions closely match the initial experimental behaviour, although the average crack arrest behaviour is best described by the 15\textdegree{} angle results. This is to be compared with the 20 to 25\textdegree{} average propagation angle measured on cross section and by tomography. Of course in real experiments, the crack angle is not constant and further work is needed (using a bifurcation criterion) to extend those prediction with a varying crack angle as a function of the crack depth. Criteria for the determination of angles of crack growth are now well established under proportional loading \cite{Erdogan1963, Sih1974, Tirosh1977}. However, fretting conditions induces non-proportional loading at crack tips which are characterized by a $K_I/K_{II}$ ratio varying during the cycle. Only a few theorical studies are devoted to non-proportional conditions and some of them \cite{Bold1992, Smith1983} showed the inapplicability of certain classical criteria to describe crack propagation. For instance, Hourlier and Pineau applied a criteria under non-proportional conditions where the crack path corresponds to the direction where the mode I fatigue growth rate is maximum \cite{Hourlier1982}. Bower suggested that the crack would tend to propagate in the direction of the maximum value of $\Delta \sigma^{max}$ \cite{Bower1988}. Lamacq showed that for fretting fatigue conditions, the $K_{II}=0$ approach worked well to predict the crack angle \cite{Lamacq1996}. Yet, there is no general agreement on which criterion is to work in the pure fretting case.

\section{Conclusion}
This paper presented a numerical framework suitable to study crack propagation under pure fretting conditions. The FEM simulations have been coupled with LEFM to successfully describe some experimental characteristic features of fretting cracks like the influence of the crack angle and the crack arrest depth. Given the highly multi-axial stress field developing under the contact, the path taken by the crack appears critical with respect to its propagation rate.
Regarding mode I, the maximum stress intensity value decreases as the crack grows and it was shown that the decrease is stronger when the crack is inclined towards the center of the contact zone. For mode II, although the stress intensity level $K_{II}$ remains high, its range $\Delta K_{II}$ progressively decrease to zero if the friction of crack faces is taken into account, this ultimately being the cause of the crack arrest. If crack propagation is assume to occur without friction, predicted mode II stress intensity factor ranges are much higher which leads to longer fretting cracks.

Two directions appear particularly important to extend this work. First, the effect of crack bifurcation should be investigated. Many criteria are available from the literature to try to predict the path of a crack in a multiaxial stress field but the non-proportional nature of fretting loads should be accounted for. Second, the present model cannot yet account for microstructural effect on the crack path as shown by a number of studies. Cyclic plastic deformation at the crack tip should be investigated using a suitable material constitutive behaviour such as crystal plasticity.

\section*{Acknowledgments}
The authors are grateful to Professor G. Cailletaud for fruitful discussion and comments.

\bibliography{bibliography}

\end{document}